\documentclass[twocolumn]{jpsj3}
\usepackage{txfonts}
\usepackage{subfigure}

\title{Vacuum-fluctuation-induced Dephasing of a Qubit in 
 Circuit Quantum Electrodynamics}

\author{Y.-W. Kim$^1$, K.-H. Lee$^{1,2}$, 
 Kicheon Kang$^{1,3}$\thanks{kicheon.kang@gmail.com}}
\inst{$^1$Department of Physics, Chonnam National University, Gwangju 500-757, Korea \\
$^2$Department of Physics, Ben Gurion University of the Negev, Beersheba 84105, Israel \\
$^3$Asia Pacific Center for Theoretical Physics, POSTECH, Gyeongbuk 790-784, Korea
}
\abst{We investigate the measurement-induced dephasing of a qubit coupled 
with a single-mode cavity in the vacuum limit.
Dephasing of the qubit state takes place through the entanglement
of the qubit and the single probe photon sent to the cavity, while the
cavity mode never occupies a photon. We find that
the qubit state is dephased even if the cavity is always in
the vacuum state. This dephasing is caused purely by the interaction
between the qubit and the vacuum field. We also show that this
vacuum-fluctuation-induced dephasing takes place much faster than the
spontaneous decay of the qubit excited state, and therefore our prediction
is observable in a real experiment.}

\pagebreak

\begin{document}
\maketitle

{\em Introduction} - 
The complementarity principle~\cite{Bohr83} is the basis of quantum physics.
It demands that a possibility of extracting the state information
of a ``system" with an ``apparatus" inevitably causes a reduction in the
quantum interference of the state being monitored.
This so-called ``measurement-induced dephasing" has been clearly confirmed
in real experiments with photons~\cite{Zou91}, Rydberg atoms~\cite{Brune96},
and electrons in solid state interferometers~\cite{Buks98,chang08}.
Circuit (cavity) quantum electrodynamics~(QED)
architecture~\cite{Blais04,Schoelkopf08,You2011} is an ideal
playground for studying
complementarity and measurement-induced dephasing. The dispersive measurement
scheme provides a non-invasive readout suitable for quantum information
processing, and is also ideal for testing the complementarity principle.
In this system, it is generally believed that the
measurement-induced dephasing is caused by the photon number
fluctuations in the
cavity~\cite{Blais04,Gambetta06,Boisonneault09,Serban07,Gambetta08,Sears12},
as it has been experimentally
demonstrated~\cite{Bertet05,Schuster05}.

In contrast to this semiclassical photon shot-noise dephasing,
here we propose that measurement-induced dephasing takes place
even in the vacuum limit of the cavity where the photon number fluctuation is
completely suppressed. To be specific, we consider a qubit coupled to a
resonator with a single photon mode in the dispersive limit.
We show that the qubit is dephased due to the vacuum field in the resonator
with the condition that the photon number is always zero in the cavity.
This demonstrates a novel property of the vacuum manifested in the context
of the quantum mechanical complementarity. 

%
{\em Model Hamiltonian} - 
A qubit interacting with a single-mode resonator is represented by
the Hamiltonian,
\begin{subequations}
\begin{equation}
 H = H_0 + H_c + V_{0c} \,,
\end{equation}
where $H_0$ denotes the qubit+cavity mode represented by
the usual Jaynes-Cummings model~\cite{Jaynes63} with rotating wave
approximation. $H_c$ and $V_{0c}$ represent the continuum modes outside
the cavity and the cavity-continuum interaction, respectively:
\begin{eqnarray}
 H_0 &=& \hbar\omega_r \left(a^\dagger a + \frac{1}{2} \right)
      + \frac{\hbar\omega_q}{2}\hat{\sigma}_z
      + \hbar g(a^\dagger\sigma^- + \sigma^+ a), \\
 H_c &=& \sum_k \hbar\nu_k \left( a_k^\dagger a_k + \frac{1}{2} \right), \\
 V_{0c} &=& \sum_k \hbar\alpha_k \left( a_k^\dagger a + a^\dagger a_k \right).
\end{eqnarray}
\end{subequations}
The parameters $\omega_r$, $\omega_q$, $g$, $\nu_k$, and $\alpha_k$ denote the resonance
frequency
of the cavity, qubit frequency, the coupling between the qubit and the cavity,
the continuum mode frequency, and the coupling between the cavity and the continuum,
respectively. There are two important parameters governing the characteristics
of the system in our context: the detuning of the qubit and
the cavity frequencies, $\Delta\equiv \omega_q - \omega_r$, and the
decay rate of the cavity mode, $\Gamma\equiv 2\pi\sum_k|\alpha_k|^2
\delta(\omega_r - \nu_k)$, assumed to be independent of the frequency.
In the dispersive limit ($|\Delta|\gg g$), $H_0$ is transformed to an
effective Hamiltonian,
\begin{equation}
\tilde{H}_0 = \hbar \omega_r \left(a^\dagger a + \frac{1}{2} \right)
    + \frac{\hbar}{2} \left[
      \omega_q + 2\chi \left( a^\dagger a + \frac{1}{2} \right)
                     \right] \hat{\sigma}_z \,,
\label{DispHamil}
\end{equation}
where $\chi = g^2/\Delta$ is the dispersive shift.

%
{\em Measurement-induced dephasing in the vacuum limit of the resonator} -
The measurement-induced dephasing is generally regarded as a result
of the photon number fluctuations. For instance, in a theory
based on the continuous weak measurement~($\chi\ll\Gamma$)~\cite{Blais04,Gambetta06},
the phase factor of the qubit decays as
 $\langle e^{i\varphi(t)} \rangle = \exp{(-\Gamma_\phi t)}$, with the
dephasing rate
\begin{equation}
 \Gamma_\phi = 2\theta_0^2 \Gamma \bar{n} \,,
\label{eq:dephasing-number}
\end{equation}
where $\theta_0\equiv 2\chi/\Gamma$ and $\bar{n}$ is the average
photon number in the cavity.
%
This property of dephasing rate, which is governed by the photon number
fluctuation, has been confirmed in
experiments~\cite{Bertet05,Schuster05}.
The photon shot noise dephasing is verified also in the strong measurement
limit where single photons fully dephase the qubit~\cite{Sears12}.
Both in the weak and in the strong measurement limits,
the photon number fluctuation in the cavity has been regarded
as the source of the measurement-induced dephasing. In these
studies, the effect of the vacuum field was not taken into account,
and the dephasing rate vanishes in the vacuum limit.

\begin{figure}[!htp]
\centering
\includegraphics[width=80mm]{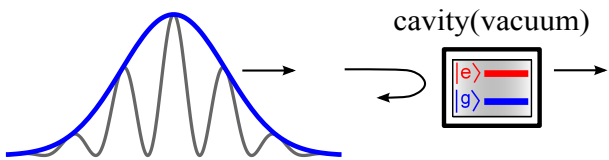}
\caption{Schematic diagram of a qubit embedded in a cavity.
Single ``probe" photons are sent to the cavity,
but the cavity is always in the vacuum state during the interaction process.
}
\label{Setup}
\end{figure}

%
We investigate the role of the vacuum fluctuation in the
measurement-induced dephasing.
Our aim is to show that $\Gamma_\phi$ does not vanish even when
the cavity is always in the vacuum state.
Let us consider a measurement setup~(see Fig.~\ref{Setup}) where single ``probe" photons
are sent to a cavity interacting with a qubit.
We consider a circuit QED system for a possible realization, but a cavity
QED setup can also be used. The key point is that, during its travel through
the cavity, the outcome of a probe photon becomes entangled with the qubit state,
and thereby produces a ``measurement-induced dephasing" of the qubit.

The initial state of the qubit-probe photon composite,
$$ |\psi(0)\rangle = (c_g|g\rangle + c_e|e\rangle) \otimes |D\rangle \,,$$
evolves into
\begin{equation}
 |\psi(t)\rangle  = c_g|g\rangle\otimes|D_g(t)\rangle
           + c_e|e\rangle\otimes|D_e(t)\rangle \,.
\end{equation}
The time evolution of a probe photon state, $|D_s\rangle$,
depends on the qubit state $s$ ($=g$ or $e$), and is given by
\begin{equation}
 |D_s(t) \rangle = A_s(t)|1,0\rangle
    + \int B_s(k,t) |0,1_k\rangle\, d k ,
\label{DetectorStateVector}
\end{equation}
where the photon state is given in the number state representation
$|n,n_k\rangle$
with $n$ and $n_k$ being the photon numbers in the cavity and in the mode $k$
of continuum, respectively.

First, we consider a simple case of injecting a monochromatic single photon,
$|0,1_{k_0} \rangle$.
This case is specified by the initial condition,
\begin{equation}
A_s(0)=0\,,\quad B_s(k,0) = \delta(k-k_0) \,.
\end{equation}
The time evolutions of the coefficients $A_s(t) $ and $B_s(k,t)$ are found to be
\begin{subequations}
\begin{eqnarray}
 A_s(t) &=& \frac{\alpha_{k_0}}{\omega_s -\nu_{k_0} - i\Gamma/2}
    \left( e^{-\Gamma t/2} - e^{i(\omega_s -\nu_{k_0})t} \right),
\label{At} \\
 B_s(k,t) &=& \delta(k-k_0)
  + \frac{\alpha_{k}^* \alpha_{k_0}}{\omega_s-\nu_{k_0} - i\Gamma/2}
\label{Bkt} \\
  &\times& \left(
  \frac{e^{-i(\omega_s-\nu_k- i\Gamma/2)t}-1}{
       \omega_s - \nu_k - i\Gamma/2}
  - \frac{e^{-i(\nu_{k_0} - \nu_{k})t} - 1}{\nu_{k_0} - \nu_k }
       \right),
\notag
\end{eqnarray}
\end{subequations}
where $\omega_s = \omega_r\pm\chi$ (for $s=e/g$) is the qubit-dependent
cavity frequency shifted by the dispersive interaction.
The qubit state information is eventually encoded in the asymptotic limit
of the photon state $|D_s(t\rightarrow\infty)\rangle$. This asymptotic limit,
obtained by taking $\nu_k\rightarrow \nu_k+i\epsilon$
and $t\rightarrow\infty$, is given by
\begin{equation}
 |D_s(\infty) \rangle =
    \frac{\omega_s-\nu_{k_0}}{\omega_s-\nu_{k_0}-i\Gamma/2}
    \left| 0, k_0 \right\rangle
  + \frac{i\Gamma/2}{\omega_s - \nu_{k_0} - i\Gamma/2}|0,-k_0\rangle . \label{AsymptDetector}
\end{equation}
The coherence factor, defined as
\begin{equation}
\lambda \equiv \langle D_e(\infty)|D_g(\infty)\rangle \,,
\label{eq:lambda}
\end{equation}
is a measure of interference between the two qubit states. 
For calculating $\lambda$, we use the condition $\nu_{k0}=\omega_g$, 
that is, the input frequency is in resonance with the dressed cavity frequency
when the qubit is in the ground state. In fact, this is an optimal measurement
condition, and we find 
\begin{equation}
\lambda = \frac{i\Gamma/2}{2\chi + i\Gamma/2} \,. \label{MonoLambda}
\end{equation}
In the strong measurement limit ($\chi\gg\Gamma$), $\lambda\approx0$.
This implies that the qubit state is fully dephased when a single photon is sent
to the cavity. It should be noted that this dephasing has nothing to do
with the photon number fluctuation, but originates purely from
the vacuum field in the cavity.
To confirm this statement, we have to check the occupation probability of 
a photon in the cavity at an arbitrary time $t$, $P_1(t)$, which is given by
\begin{equation}
P_1(t) = |A_s(t)|^2,
\end{equation}
in our case. 
As one can find from Eq.~(\ref{At}),
$A_s(t)\sim\alpha_{k_0} /\Gamma$. Note that $\Gamma=2\pi|\alpha|^2\rho$, where
$\rho$, the density of states at the input frequency $\nu_{k0}$, 
is a macroscopic quantity and is proportional
to $M$, the number of modes in the reservoir which are coupled to the cavity.  
Therefore, $\alpha_{k0}
\propto 1/\sqrt{M}$, and we find that $P_1(t)$ is proportional to $1/M$.
Since $M\rightarrow\infty$ for a reservoir with continuous spectrum, $P_1(t)=0$.
This implies that photon number fluctuation is absent in the cavity,
since a photon never occupies the cavity mode during the interaction process.
The full dephasing of
the qubit in the strong measurement limit is understood only in terms
of the interaction between the qubit and the vacuum field, which transcends
 the semiclassical concept of dephasing generated by the
photon number fluctuation.

%
The above discussion with a monochromatic probe photon
already includes the essence of the vacuum-fluctuation-induced
dephasing.
A drawback of this treatment is that the dephasing time is infinity
in this ideal case, because the asymptotic limit cannot be reached in a
finite time interval with a monochromatic wave.
Next, we treat a more realistic case of
sending a single photon with its frequency dispersion $\Delta\nu$.
This can be modeled, for example, by an input of the Gaussian wave packet:
\begin{equation}
 |D_s(t=0)\rangle =
    N_0 \int e^{-(\nu_k-\nu_{k_0})^2/(2\Delta\nu)^2} |0,1_k\rangle\, d\nu_k .
\label{input-packet}
\end{equation}
The time evolution of this state is given by $|D_s(t)\rangle =
\exp{(-iHt)}|D_s(0)\rangle$, and calculating the
coherence factor (Eq.~(\ref{eq:lambda})) from the asymptotic limit of
$|D_s(t)\rangle$ is straightforward. We find
\begin{subequations} \label{eq:GaussLambda}
\begin{equation}
 \lambda = \lambda\left( \frac{2\Delta \nu}{\Gamma},
           \frac{2\chi}{\Gamma} \right),
\end{equation}
where the function $\lambda(x,y)$ is given by
\begin{align}
 \lambda(x,y) =& 1-\sqrt{\frac{\pi}{2}} \frac{y}{x(y+i)}
    \Bigg[ e^{(1-i2y)^2/2x^2} \mathrm{erfc}\left( \frac{1-i2y}{\sqrt{2}x} \right) \notag \\
           &+ e^{1/2x^2} \mathrm{erfc}\left(\frac{1}{\sqrt{2}{x}}\right)
    \Bigg] \,.
\end{align}
\end{subequations}

The probe photon reaches the asymptotic limit
more rapidly as the pulse length ($\Delta x$) is reduced, or equivalently,
as $\Delta\nu$ increases~(see Fig.~2). This implies
that a measurement can be performed faster for a larger dispersion in the input
frequency.
For a pulse that is too short, however, the probe photon cannot
really detect the qubit state ($|\lambda|\sim1$) because of the large frequency
dispersion $\Delta\nu$
that obscures the resolution in the measurement.
This is verified in the behavior of $|\lambda|$
as a function of $\Delta\nu$ (Fig.~2(b)).
The optimal measurement can be obtained by imposing
the condition $|\lambda|=1/e$. The dephasing time 
is determined by
the travel time of the input wave through the cavity that satisfies
the condition $|\lambda|=1/e$.
In the strong measurement limit, $\chi\gg\Gamma$,
we find that the vacuum-induced dephasing rate, $\Gamma_\phi$, is
proportional to the cavity decay rate $\Gamma$ as
\begin{equation}
 \Gamma_\phi \simeq 1.05\Gamma \,.
 \label{eq:SM-Rate}
\end{equation}

Our discussion of the dephasing rate can be easily extended to the case
where the cavity has $n$ photons. In this case, the effective coupling of
the cavity and the reservoir is enhanced by the factor of $n+1$ compared
to the vacuum limit, and thus
the dephasing rate is also enhanced by the same factor as
\begin{equation}
 \Gamma_\phi(n) =(n+1)\Gamma_\phi \,.
\end{equation}

%
The vacuum-fluctuation-induced dephasing takes place not only in the strong
measurement limit, but also in the weak measurement limit ($\chi \ll \Gamma$).
In this case, the single probe photon only has a negligible effect on the
qubit state. Instead, the qubit state can be dephased by a continuous
injection of probe photons. The interference of the qubit state is reduced
by the factor $|\lambda^m|$ in this case, which can be written as
\begin{equation}
 |\lambda^m| \simeq e^{-\Gamma_\phi m\delta t}, 
\end{equation}
where $m$ ($\gg 1$) and $\delta t$ are the number of injected photons
and the time interval between the successive transmission of the input packets,
respectively.
$\lambda$ is the coherence factor due to a single probe photon, given by
Eq.\eqref{eq:GaussLambda}.
In general, the vacuum-fluctuation-induced dephasing rate, $\Gamma_\phi$,
depends on the packet width of the input photons.
An optimal setting of measurement is the case where the
frequency dispersion ($\Delta\nu$) of the input wave is comparable to the cavity-reservoir
coupling ($\Gamma$).
We find that, for $\Delta\nu=\Gamma$,
\begin{equation}
 \Gamma_\phi = \beta \frac{(2\chi)^2}{\Gamma} \, ,
 \label{eq:CWM-Rate}
\end{equation}
with the constant $\beta = 2(3- 2e\sqrt{\pi}\,\mathrm{erfc{[1]}}) = 2.97$.

The behavior of the vacuum-fluctuation-induced dephasing rate is
summarized in Fig.~4. In the strong measurement limit, 
$\Gamma_\phi$ is determined from the condition $|\lambda|=1/e$ with $\lambda$
given by Eq.~\eqref{eq:GaussLambda}. The result for a continuous weak measurement
limit is obtained from Eq.~\eqref{eq:CWM-Rate}.

%
A necessary condition for an experimental observation of the
vacuum-field-induced dephasing
is that the dephasing should take place before the spontaneous relaxation of the
qubit state. That is, the dephasing rate should be larger than the rate of
spontaneous relaxation.
Here we show that this is
indeed the case in both the strong and the continuous weak measurement
limits.
The spontaneous decay rate in our system,
denoted by $\gamma$, is governed by the Purcell effect~\cite{Purcell46},
and is given as
\begin{equation}
 \gamma = 2g^2 \frac{\Gamma/2}{\Delta^2+(\Gamma/2)^2} \,.
\end{equation}
We consider only the dispersive limit in which the detuning $\Delta$ is
greater than the coupling strengths $g$ and $\Gamma$. In this limit,
$\gamma$ is reduced to
\begin{equation}
 \gamma \simeq  \left(\frac{g}{\Delta}\right)^2\Gamma \,.
 \label{eq:spontaneous}
\end{equation}
Comparing Eq.~\eqref{eq:SM-Rate} and Eq.~\eqref{eq:spontaneous},
we find that $\Gamma_\phi/\gamma\simeq (\Delta/g)^2$ in the strong measurement
limit ($\chi\gg\Gamma$).
That is, the dephasing takes place faster than the spontaneous decay by the
factor $(\Delta/g)^2$. 
In the case of continuous weak measurement,
the ratio is given by $\Gamma_\phi/\gamma\simeq 12 (g/\Gamma)^2$
(compare Eq.\eqref{eq:CWM-Rate} and Eq.\eqref{eq:spontaneous}),
implying that the dephasing can be observed in the strong coupling regime
($g\gg\Gamma$).
Therefore, in any case, our prediction of vacuum-induced dephasing is observable
before the qubit is spontaneously relaxed to its ground state.

%
Finally, we discuss in more detail the meaning of cavity vacuum
during the interaction between the qubit and the probe photons. As previously
discussed, the occupation probability of a photon in the
cavity is always zero, even if the photons pass through the cavity.
Here, a puzzling question is raised: How can a photon pass through the cavity
if the probability of finding it in the cavity is always zero?
This appears to be paradoxical because the photons are transmitted and reflected
only via the cavity. The answer to this question is that, in fact, this
classical intuition fails. The paradox is based on the semiclassical
picture of a
sequential process that a photon is first captured in, and then escapes from,
the cavity. The reasoning based on this picture would lead to the conclusion that,
if the probability of the photon being captured is zero,
there would be no chance of transmitting a photon.
Interestingly, this semiclassical concept is invalid, and we find
a finite probability
of finding a photon transmission through the cavity while it is never
occupied, even virtually,  in the cavity.
This is a clear example demonstrating
the counter-intuitive nature of quantum theory.
Therefore, we can conclude that the measurement-induced dephasing
investigated here originates
purely from the vacuum fluctuation. The qubit is interacting with
the vacuum electric field in the cavity, and this interaction
influences the cavity mode. The
frequency of the vacuum field depends on the qubit state, thereby transferring
the qubit state information into the probe photons.

%
{\em Conclusion} - We predicted that a qubit state is dephased by the
zero-point fluctuation of the cavity mode. This vacuum-fluctuation-induced
dephasing can be verified in a real experiment of a circuit (cavity)
QED setup. 
The vacuum-induced dephasing has a pure quantum mechanical origin which cannot
be understood in terms of the photon shot noise. The role of a vacuum as
``something" in distinction from ``nothing" has been highlighted through various
interesting phenomena such as Casimir force\cite{Casimir48}, Lamb shift\cite{Lamb47},
and spontaneous emission\cite{Weisskopf30, Einstein17}, etc.
The vacuum-fluctuation-induced dephasing is another intriguing phenomenon
manifested by the vacuum electromagnetic field.

\begin{acknowledgment}
This work was supported by the National Research Foundation of Korea under
Grant No.~2012R1A1A2003957.
\end{acknowledgment}


\begin{figure}[!htp]
\centering
	\subfigure{\label{StrongFiga}}
		\includegraphics[width=65mm]{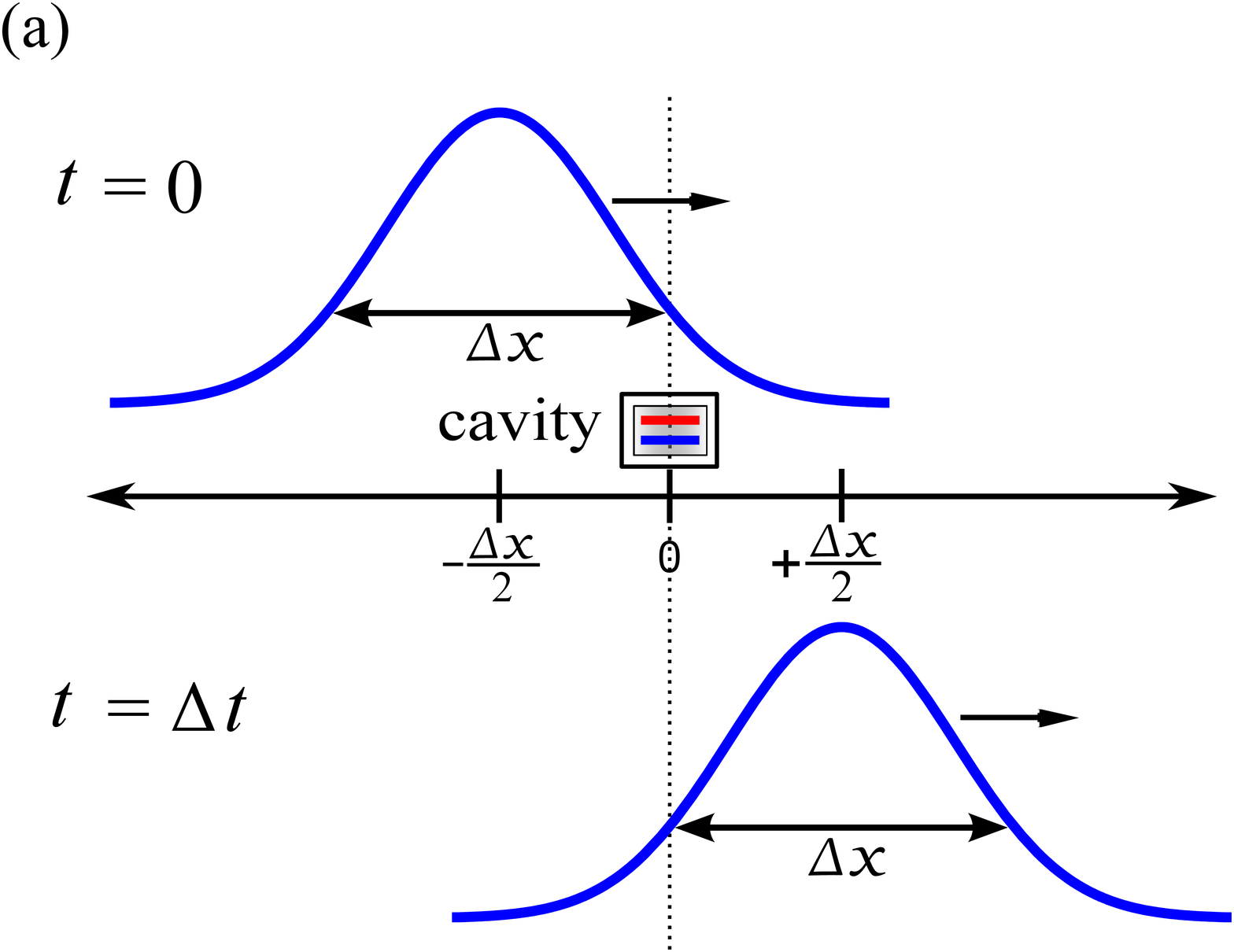} \\
	\subfigure{\label{StrongFigb}}
		\includegraphics[width=65mm]{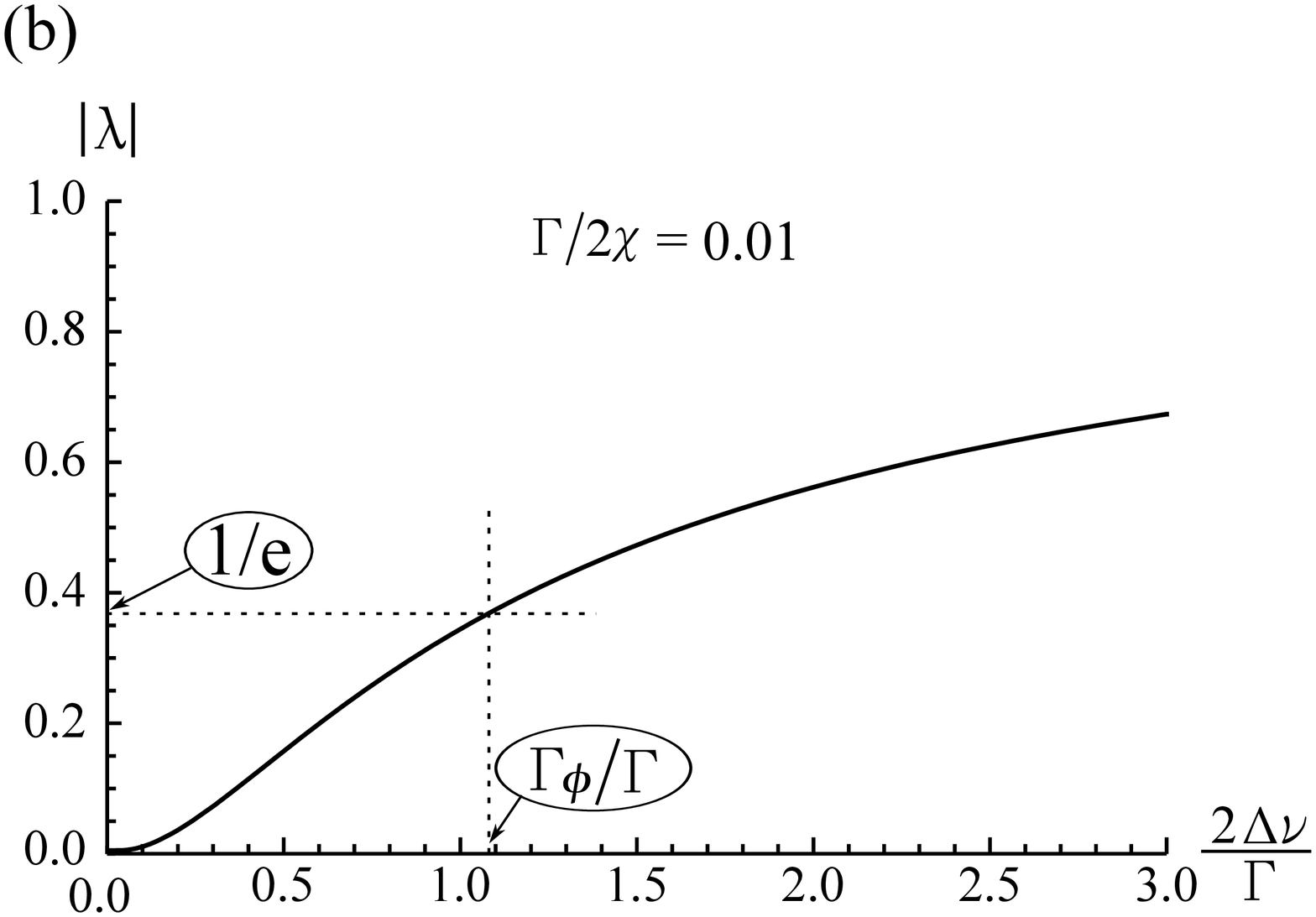}
\caption{(a) Illustration of a ``measurement"  using a single probe
 photon with the packet size $\Delta x$.
 The probe photon interacts with the cavity during the time interval of
 $\Delta t=(2\Delta\nu)^{-1}=\Delta x/c$.
(b) The coherence factor ($|\lambda|$) as a function of the frequency
 dispersion ($\Delta\nu$) of the probe photon in a strong measurement limit,
 $\Gamma/2\chi=0.01$.
 The dephasing time is determined by the travelling time ($\Delta t$)
 of the packet time together with the condition $|\lambda|=1/e$
 (see text).
}
\label{StrongFig}
\end{figure}
\begin{figure}[!htp]
\centering
\includegraphics[width=65mm]{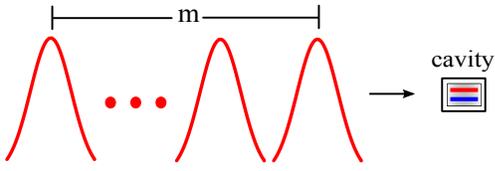}
\caption{Continuous weak measurement scheme. The measurement-induced
 dephasing is determined by a continuous injection of the probe photons,
 while a single photon dephases the qubit state only very weakly.
}
\label{CWMFig}
\end{figure}
\begin{figure}[!htp]
\centering
\includegraphics[width=70mm]{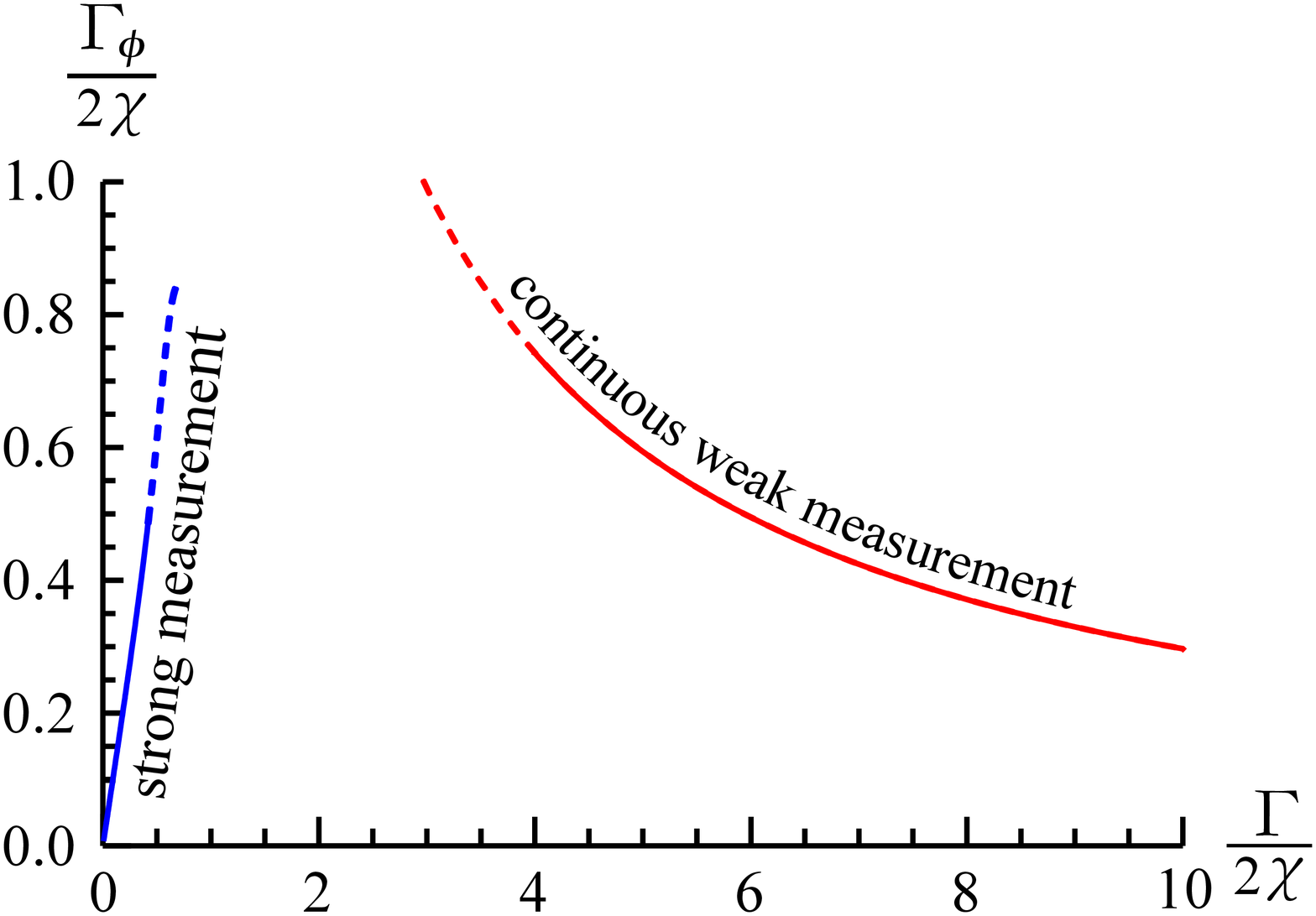}
\caption{Numerical result of the vacuum-fluctuation-induced dephasing rate
 for the strong measurement (blue line), and for the continuous
 weak measurement (red line) limits, respectively.
 }
 \label{DephasingRageFig}
\end{figure}

\end{document}